\documentclass[12pt]{iopart}

\usepackage{amsfonts,amssymb,graphics,graphicx,epsfig,color,times}

\graphicspath{{images/}}
 
\DeclareMathSymbol{\theta}{\mathord}{letters}{"23}
\DeclareMathSymbol{\rho}{\mathord}{letters}{"25}
\DeclareMathSymbol{\phi}{\mathord}{letters}{"27}
 
\DeclareMathSymbol{\vartheta}{\mathord}{letters}{"12}
\DeclareMathSymbol{\varphi}{\mathord}{letters}{"1E}
\DeclareMathSymbol{\varrho}{\mathord}{letters}{"1A}
\begin{document}

 
\newcommand{\nn}{{\mathbbm{N}}}
\newcommand{\rr}{{\mathbbm{R}}}
\newcommand{\cc}{{\mathbbm{C}}}
\newcommand{\id}{{\sf 1 \hspace{-0.3ex} \rule{0.1ex}{1.52ex}\rule[-.01ex]{0.3ex}{0.1ex}}}
\newcommand{\me}{\mathrm{e}}
\newcommand{\mi}{\mathrm{i}}
\newcommand{\md}{\mathrm{d}}
\renewcommand{\vec}[1]{\rm{\boldmath$#1$}}

\newcommand{\rB}{{\rm B}}
\newcommand{\rF}{{\rm F}}
\newcommand{\rDD}{{\rm 3D}}
\newcommand{\rED}{{\rm 1D}}

\newcommand{\cH}{\mathcal{H}}
\newcommand{\cO}{\mathcal{O}}
\newcommand{\cI}{\mathcal{I}}
\newcommand{\cZ}{\mathcal{Z}}

\newcommand{\hP}{\hat{\Psi}}
\newcommand{\hPd}{\hat{\Psi}^\dagger}
\newcommand{\hPe}{\hat{P}}
\newcommand{\hH}{\hat{H}}
\newcommand{\hA}{\hat{A}}
\newcommand{\hB}{\hat{B}}
\newcommand{\hX}{\hat{X}}
\newcommand{\hL}{\hat{L}}
\newcommand{\ha}{\hat{a}}
\newcommand{\had}{\hat{a}^\dagger}
\newcommand{\hc}{\hat{c}}
\newcommand{\hcd}{\hat{c}^\dagger}
\newcommand{\hr}{\hat{\rho}}
\newcommand{\hN}{\hat{N}}
\newcommand{\hn}{\hat{n}}
\newcommand{\hU}{\hat{U}}
\newcommand{\hV}{\hat{V}}
\newcommand{\hE}{\hat{E}}
\newcommand{\hS}{\hat{S}}
\newcommand{\hO}{\hat{O}}
\newcommand{\hcI}{\hat{\cI}}

\newcommand{\brX}{{\bf r}}
\newcommand{\bv}{{\bf v}}
\newcommand{\bt}{{\bf t}}
\newcommand{\be}{{\bf e}}
\newcommand{\bF}{{\bf F}}

\newcommand{\scl}{\hspace{-0.6cm}}
\newcommand{\scls}{\hspace{-0.3cm}}

\newcommand{\bra}[1]{\langle #1 \vert}
\newcommand{\ket}[1]{\vert #1 \rangle}
\newcommand{\braket}[2]{\langle #1 \vert #2 \rangle}
\newcommand{\ko}[2]{\left[ #1, #2 \right]}
\newcommand{\ako}[2]{\left\{ #1, #2 \right\}}
\newcommand{\expv}[1]{\langle #1 \rangle}
\newcommand{\set}[1]{\left\{ #1 \right\}}

\newcommand{\hOg}{\hat{O}_{g}}
\newcommand{\TrX}{\mathrm{Tr}}
\newcommand{\eps}{\varepsilon}

\renewcommand{\H}{\mathcal{H}}
\newcommand{\ad}{\hat{a}^\dagger}
\renewcommand{\a}{\hat{a}}
\newcommand{\nb}{\hat{n}}
\newcommand{\cd}{\hat{c}^\dagger}
\renewcommand{\c}{\hat{c}}
\newcommand{\nf}{\hat{m}}
\newcommand{\f}{\hat{f}}
\newcommand{\NN}[1]{\sum_{\langle#1\rangle}}
\newcommand{\cred}{\color{red}}
\newcommand{\cblue}{\color{blue}}
\newcommand{\cgreen}{\color{green}}
\newcommand{\Uh}{\frac{U}{2}}
\newcommand{\trace}[2]{\textsf{Tr}_{#1}\left[#2\right]}


\newcommand{\kb}{k_{\rm{B}}}
\newcommand{\Tc}{T_{\rm{c}}}
\newcommand{\Eia}{E_{\rm{ia}}}
\newcommand{\Ekin}{E_{\rm{kin}}}
\newcommand{\Etrap}{E_{\rm{trap}}}
\newcommand{\Hia}{\hH_{\rm{ia}}}
\newcommand{\Hkin}{\hH_{\rm{kin}}}
\newcommand{\Htrap}{\hH_{\rm{trap}}}
\newcommand{\tint}{t_{\rm{int}}}
\newcommand{\losc}{l_{\rm{osc}}}


\title[Fermionisation dynamics of a strongly interacting 1D Bose gas]{Fermionisation dynamics of a strongly interacting 1D Bose gas after an interaction quench}

\author{Dominik Muth$^1$, Bernd Schmidt$^{1, 2}$ and Michael Fleischhauer$^1$}
\address{$^1$ Fachbereich Physik und Forschungszentrum OPTIMAS, Technische Universit\"at Kaiserslautern, D-67663 Kaiserslautern, Germany}
\address{$^2$ Institut f\"ur Theoretische Physik, Johann Wolfgang Goethe-Universit\"{a}t, D-60438 Frankfurt am Main, Germany}

\begin{abstract}
We study the dynamics of a one-dimensional Bose gas after a sudden change of the
interaction strength from zero to a finite value using the numerical time-evolving block decimation
(TEBD) algorithm. It is shown that despite the integrability of the system, local quantities such as the
two-particle correlation $g^{(2)}(x,x)$ attain steady state values in a short characteristic time inversely
proportional to the Tonks parameter $\gamma$ and the square of the density. 
The asymptotic values are very close to those of a finite temperature
grand canonical ensemble with a local temperature corresponding to initial energy and
density. Non-local density-density correlations on the other hand approach a steady state
on a much larger time scale determined by the finite propagation velocity of oscillatory correlation waves. 
\end{abstract}

\pacs{03.75.Kk, 67.85.De, 05.30.Jp, 05.70.Ln}

 
\date{\today}
 
\maketitle

\section{Introduction}

The dynamics of interacting quantum systems from an initial non-equilibrium
state constitutes a major challenge for many-body theory. In particular the
question of thermalisation of integrable models regained
attention recently due to the experimental progress 
in ultra-cold gases. As demonstrated in a beautiful experiment by Kinoshita
et al. \cite{Kinoshita2006} for the example of a 1D ultra-cold Bose gas,
integrable systems do not thermalise in the usual sense, i.e., reduced
density-matrices relax on considerably different
time scales than in the absence of integrability. 

The speciality of the relaxation dynamics of integrable systems has been attributed to the presence of 
an infinite set of constants of motion with local character, i.e., which
can be written as sums of operators acting only over a finite spatial range. 
Although thermalisation has been studied in a large body of theoretical papers it remains a largely unsolved problem. 
Most studies of specific models have been done either for non-interacting particles 
\cite{Cramer2008} or systems that can directly be mapped to free systems such as hard-core bosons 
\cite{Rigol2007}, the Luttinger model \cite{Luttinger1963},
or the $1/r$ fermionic Hubbard model \cite{Kollar2008}.

In the present letter we analyse the dynamics of a 1D Bose gas with $s$-wave scattering
interactions, described by the Lieb-Liniger (LL) model,
after a sudden quench of the interaction strength from zero to a finite value, covering the 
full range from weak to strong interactions. Performing numerical simulations using the time 
evolving block decimation algorithm (TEBD) \cite{Vidal2003,Vidal2004}, we show that local 
quantities, in particular the local two-particle correlation $g^{(2)}(0,0;t)$, do approach 
steady-state values on a short time scale determined only by the Tonks parameter $\gamma$ 
and the particle density $\rho$. This shows that although non-local quantities such as the
momentum distribution do not approach a steady state over long times \cite{Mazets2009},
there is an equilibration in a local sense. Furthermore the asymptotic values of $g^{(2)}(x,x)$ are 
very close to those obtained from a thermal Gibbs ensemble
\cite{Yang1969}, with temperature and chemical potential determined by the initial conditions and 
the amplitude of the interaction quench. Thus it is possible to define local temperature and
chemical potential and the influence of constants of motion other than total energy and particle 
number is very small, if present at all.
Non-local quantities such as the density-density correlation approach a
steady-state distribution on a larger time scale by way of correlation 
waves propagating out of the sample.

\section{LL model and lattice approximation}

A Bose gas in one spatial dimension is described by the Hamiltonian 
\begin{equation}
 \hH = \int \!\!{\rm d}x\ \biggl[\hPd(x) \left(-\frac12\partial_x^2\right) \hP(x) + \frac{g}{2}\left.\hPd\right.^2(x)\hP^2(x) + \hPd(x) V(x)\hP(x)\biggr]
\label{eq:H}
\end{equation}
in units were $\hbar=m=\kb = 1$. Here $\hat \Psi(x)$ is the field operator of the Bose gas in
second quantisation, $V(x)$ some possible trap potential, and $g$ the strength of the local particle-particle
interaction.
The latter is characterised by the dimensionless Tonks parameter
$ \gamma = {g}/{\rho}$, where $\rho=\expv{\hPd(x)\hP(x)}$ is the 1D density of particles. 
Specifically we consider here a system initially prepared in the non-interacting ground state.

The initial {\it canonical} 
state has locally only diagonal elements. In the course of interactions
non-diagonal elements are not created. Thus the reduced local density matrix 
is entirely determined by the number distribution, and the quantities of interest are the density
$\varrho$ and the local 
two-particle correlation
$g^{(2)}(x, x,t)$, where
\begin{equation}
 g^{(2)}(x,y,t) = \frac{\expv{\hPd (x)\hPd (y)\hP(y)\hP(x)}}{\rho(x)\rho(y)}.
\end{equation}
In principle also the corresponding higher-order moments are nonzero. They will, however, not be
considered here.

We study the dynamics after the interaction quench numerically by means of the 
the time-evolving block decimation algorithm \cite{Vidal2003,Vidal2004}.
As for any numerical methods this requires the use of a discretised version of the
Lib-Liniger model.
The discretization of the LL model leads to the
(non-integrable) Bose-Hubbard model \cite{Schmidt2007}
\begin{equation}
\hH = -J\sum_{j} (\ha_j^\dagger \ha_{j+1} +h.a.) + \frac{U}{2}\sum_j \ha_j^{\dagger\, 2}\ha_j^2 + D\sum_j\ha_j^\dagger\ha_j,\quad x_j= j\, \Delta x.
\label{eq:Hdiscrete}
\end{equation}
Here $J=1/(2\Delta x^2)$, $U=g/\Delta x$, and $D=1/\Delta x^2$, with $\Delta x$ being the lattice constant of the discretization grid. 
The appropriateness of discretised lattice models to describe continuous interacting Bose or Fermi gases
in the limit $\Delta x \to 0$ has been discussed and verified in a number of earlier papers
\cite{Schmidt2007,Muth2010}. Note furthermore that using the
boson-fermion duality in 1D the LL model can be mapped to an integrable lattice model, the spin 1/2 XXZ model \cite{Muth2010}.
For some data sets we used both, Equation (\ref{eq:Hdiscrete}), and the XXZ lattice model to verify that in the
considered limit the non-integrability of (\ref{eq:Hdiscrete}) has no influence on the results.

\begin{figure}
   \center
a)\begin{minipage}[b]{0.7\columnwidth}
\epsfig{file=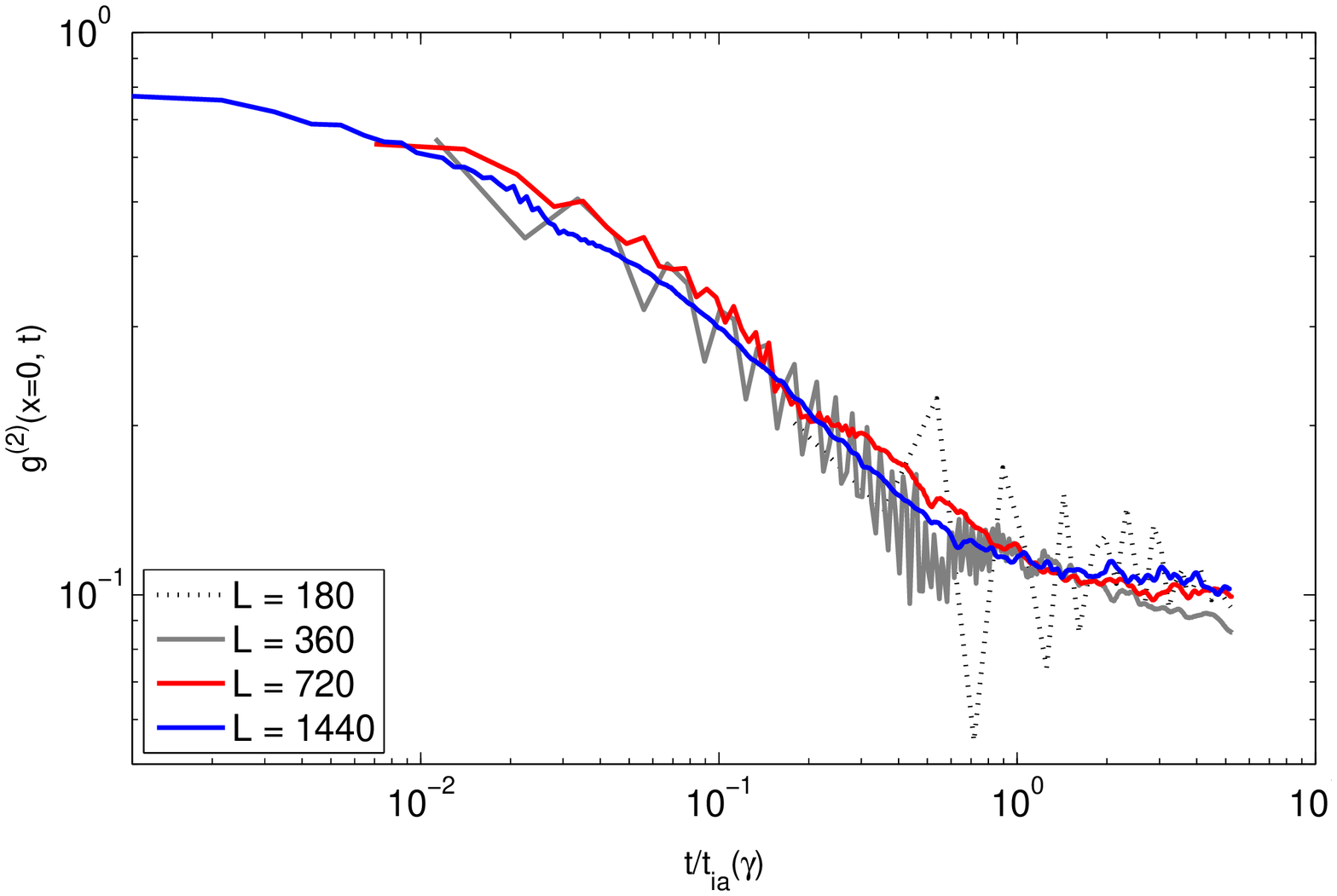,width=0.9\columnwidth}
\end{minipage}
b)\begin{minipage}[b]{0.7\columnwidth}
\epsfig{file=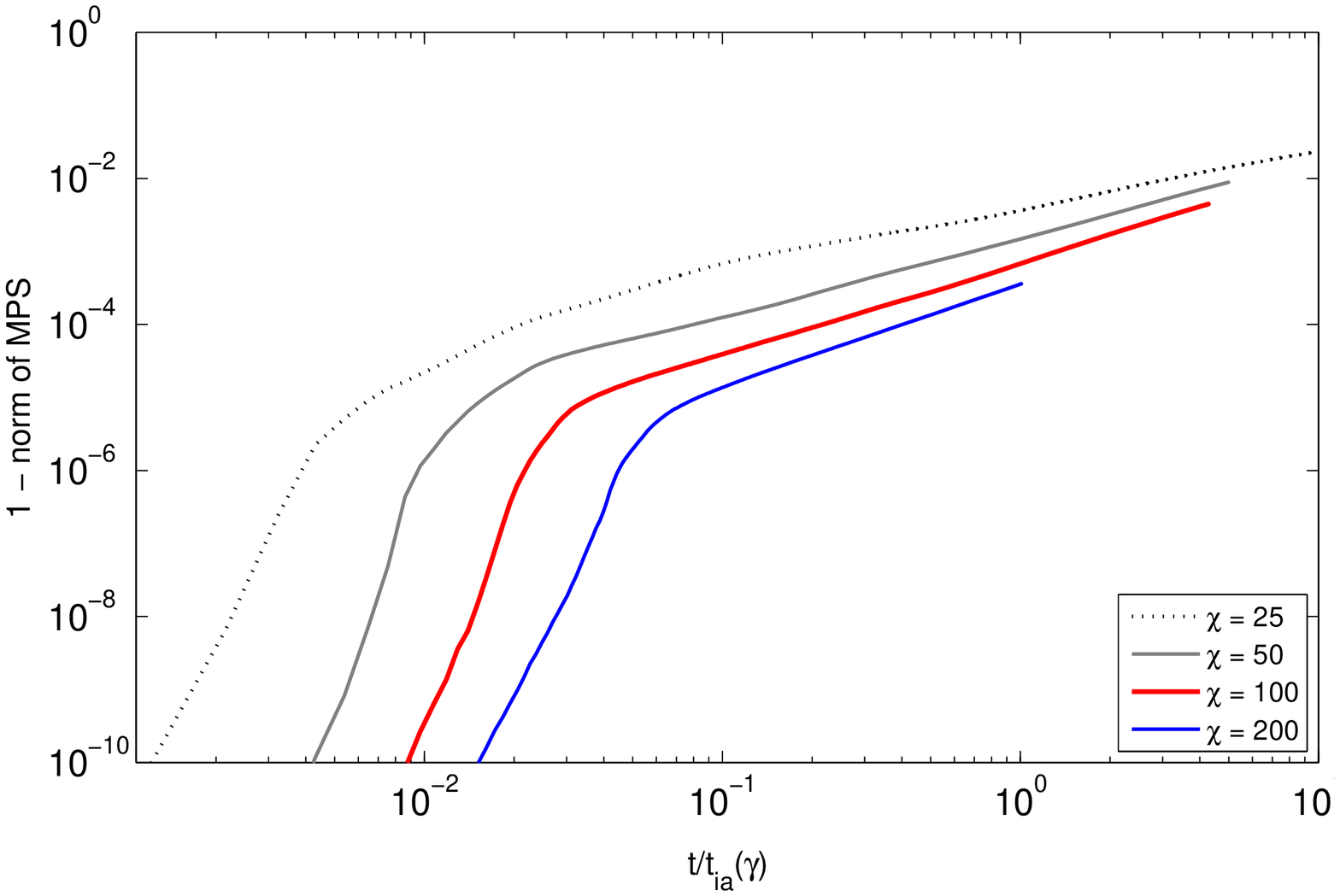,width=0.9\columnwidth}
\end{minipage}
    \caption{
{\it a) Time evolution of normalised local 
two-particle correlation $g^{(2)}(0, 0; t)$ (see section \ref{dynamics}) for 
$N=9$ particles and $\gamma = 200/9$ for increasing lattice length, corresponding to 
finer grid sizes. One clearly recognises oscillations which are lattice artifacts and
which only disappear for the largest lattice sizes. 
{\it b)} Accumulated truncation error of the norm of the MPS in the dynamical TEBD
algorithm for $\gamma = 200/9$ and increasing bond dimension
$\chi$.}
}
  \label{fig:TEBD-error}
 \end{figure}

It turned out that for the simulation of dynamics the necessary grid sizes are much smaller than
for equilibrium simulations \cite{Schmidt2007,Muth2010}. 
Empirically we found that in order to minimise lattice artifacts resulting into numerical
errors, the average number of particles per lattice site $\langle \hat n\rangle$ should
be small compared to $1/\gamma$, where $\gamma = g/\rho$ corresponds to the density at the centre of the cloud, i.e.
$\rho \to \rho(x=0)$.
This can be explained as follows: The interaction energy of a two-particle collision in the 
lattice, i.e., $g /\Delta x$ should be smaller than the bandwidth of the 
lowest Bloch band, $\sim 1/\Delta x^2$, in order not to see lattice artifacts. At the edges of the cloud where 
collisions become less probable the condition is less strong. To accommodate the requirement of a very small $\langle \hat n\rangle$ at the centre of the cloud we used space dependent grid sizes such that the average boson number per site was constant for the centre part of the particle distribution. Nevertheless to approximate the continuous model sufficiently well, very fine grids were needed leading to rather large lattice sizes on the order of up to 2880, which is rather challenging. In order to illustrate the effects of discretization we have plotted in Figure \ref{fig:TEBD-error}a the local two-particle correlation (see following section) for  $\gamma = 200/9$ and increasing
lattice sizes $L$, corresponding to finer grids. One clearly recognises oscillation artifacts which only slowly disappear with increasing $L$.

The convergence of the TEBD scheme was checked by varying the bond dimension $\chi$ of the matrix product state
(MPS) and calculating the truncation error in the state norm accumulated during the time evolution.  
In Figure \ref{fig:TEBD-error}b the accumulated truncation error is plotted for $\gamma = 200/9$ and increasing
values of $\chi$ from 25 to 200. One recognises that for the maximum value of $\chi = 200$ which we were able to
use, the truncation error is below the level of $10^{-3}$ for the time scale of interest.
This value is larger then the accuracy typically reached in ground state calculations. However we are not at the point where the cut-off explodes, which typically happens in dynamical simulations at some point. Finally the matrix dimension required to achieve a given accuracy does not depend on
the discretization length, i.e. the number of lattice sites used. It is rather the number
of particles which determines the complexity of the calculations. Thus the restriction to a moderate particle number allows us to work on lattice large compared to other applications of the algorithm.

\section{Local Dynamics}\label{dynamics}

In order to be able to perform numerical simulations with a fixed number of particles (up to 18, 
which corresponds to the experiments in references \cite{Paredes2004,Haller2009}) we have to work with a finite size system. Therefore we assumed an initial weak harmonic trapping potential
$V(x)= \frac12\omega^2x^2$. Open or periodic boundary conditions, which can also be dealt with by the TEBD algorithm, could be used alternatively. Initially the Bose gas is in the canonical ground state ($T=0$)
of non-interacting bosons in the trap, for which the matrix product representation is
analytically known, since it is a product of single particle states.
At $t=0$ we suddenly switch the interaction strength from zero to a finite
value. At the same time the trap, the only purpose of which is was the preparation of an
appropriate initial state, is switched off. On the time scales we are interested in, the
density distribution does not change, so that the presence of a trap would be
of no relevance. This also allows to apply the results of the present analysis to a homogeneous gas.

\begin{figure}
   \center\epsfig{file=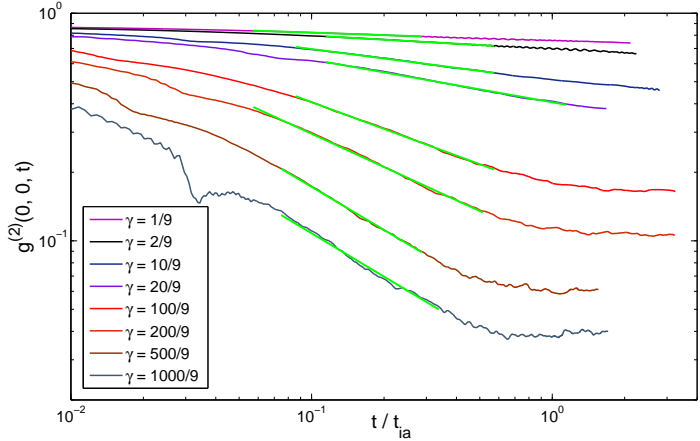,width=0.6\columnwidth}
    \caption{ (Colour online) Time evolution of normalised local 
two-particle correlation $g^{(2)}(0, 0; t)$ after a sudden switch on of interactions at $t=0$
obtained from a numerical TEBD simulation for 9 particles initially prepared in 
the non-interacting ground state of a harmonic trap. An intermediate
power-law decay with an exponent that is monotonous in $\gamma$ is apparent. The
lattice size was up to $L=2880$ for the strongest interactions corresponding to a lattice spacing of $\Delta x/l_{\rm osc}\approx 6.15
\cdot10^{-4}$ at the trap centre.}
  \label{fig:g2-local}
 \end{figure}

The initial state has a Gaussian density distribution 
$\rho(x) = N \sqrt{\frac{\omega}{\pi}}e^{-\omega x^2}$ with $l_{\rm osc}=\frac{1}{\sqrt{\omega}}$ 
being the oscillator length. 
Figure \ref{fig:g2-local} shows the time evolution of $g^{(2)}(0, 0; t)$ with time
normalised to a characteristic time scale $t_{\rm ia}$ for different values of $\gamma$. 
One recognises after an initial phase a power-law decay with an exponent that is monotonous 
in the interaction parameter. At times close to $t_{\rm ia}$ a steady
state value is attained indicating that a local equilibrium is reached.
I.e., although globally a LL gas does not thermalise \cite{Kinoshita2006}, local quantities
do. 
The time scale $t_{\rm ia}$ of the local dynamics can be estimated from Equation (\ref{eq:Hdiscrete}).
The repulsive interaction $U\hat n(\hat n-1)$ causes particle number fluctuations
to be driven out of a given lattice site. This happens in the following way: Initially all
components of the state vector have the same phase and tunnelling has no effect. 
However, due to the interaction, components
with different particle number attain a differential phase shift and 
are subsequently coupled to states in adjacent lattice sites
by tunnelling with rate $J$. Since in the limit $\Delta x \to 0$ we have $J\gg U$, the maximum
rate of this process is limited by the 
average interaction energy per particle $U \langle \hat n\rangle$. Thus we have
\begin{equation}
 t_{\rm ia} = \frac{1}{U \langle \hat n\rangle}=\frac{1}{g\rho}=\frac{1}{\gamma \rho^2}.
\end{equation}
Note that already for moderate interaction strength, $\gamma \gg 1/N^2$, this time is much shorter
as for example the oscillation time $t_{\rm osc}$ in the trap. 
\begin{equation}
 t_{\rm ia} \sim \frac{l_{\rm osc}^2 }{\gamma N^2} = t_{\rm osc} \, \frac{1}{\gamma N^2}\, \ll\, t_{\rm osc}
\end{equation}
Note furthermore that although the characteristic time of the expansion of the gas after switching
off the trap becomes much shorter than $t_{\rm osc}$ for larger interactions, we found that the
density profile did not change on the scale of $t_{\rm ia}$ even for the largest values of 
$\gamma$ used, since also $t_{\rm ia}\sim 1/\gamma$.Note furthermore that although the characteristic time of the expansion of the gas after switching
off the trap becomes much shorter than $t_{\rm osc}$ for larger interactions, it will be large compared to $t_{\rm ia}$. This is because the kinetic energy transferred to the particles will be of the order $\gamma$ and therefor their characteristic speed will be of the order $\sqrt{\gamma}$ only. Accordingly we found numerically that the density profile did not change on the timescale $t_{\rm ia}$ even for the largest values of 
$\gamma$ used.
Whether or not the thermalised local correlation will adiabatically follow the density
evolution after longer times, i.e. when the expansion of the cloud sets in, cannot
be concluded from our simulations. We would however expect
such a behaviour.

The fluctuations in the plots are artifacts of the discretization, which leads to
an oscillatory behaviour of $g^{(2)}$ on top of the continuous-system time evolution. These
artifacts, which are most pronounced for larger interactions, could not be eliminated completely 
even for the smallest grid sizes used. As a result the asymptotic values of $g^{(2)}(0,0,t)$
can only be given with a certain error.

\begin{figure}
   \center\epsfig{file=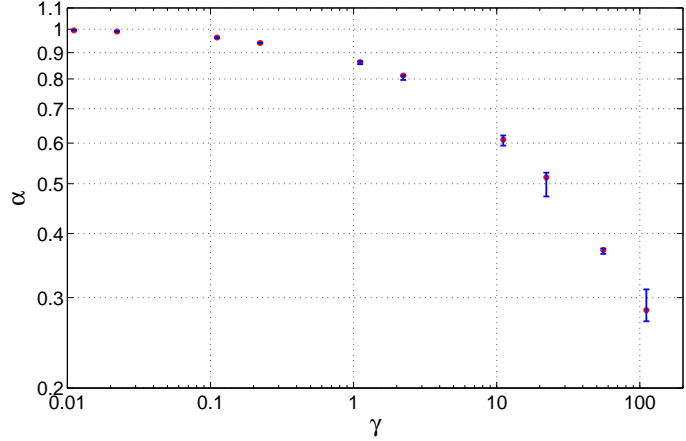,width=0.6\columnwidth}
    \caption{ (Colour online) Exponents $\alpha-1$ of the intermediate power-law decay of $g^2(0, 0; t)$ 
 in figure \ref{fig:g2-local} as function
of $\gamma$. Error-bars indicate systematic fitting error.}
  \label{fig:g2-exponent}
 \end{figure}

In figure \ref{fig:g2-exponent} we have plotted the exponents obtained from a fit
to the curves in Figure  \ref{fig:g2-local} which for intermediate times follows
a power law
\begin{equation}
 g^{(2)}(0, 0; t) \sim \left(\frac{t}{\tint}\right)^{\alpha-1}.
\end{equation}
The exponent is a monotonous function of the interaction strength and slowly approaches
the limit $-1$ for $\gamma \to \infty$, i.e., for a Tonks-Girardeau gas \cite{Chang2008}. 

\begin{figure}
   \center
\epsfig{file=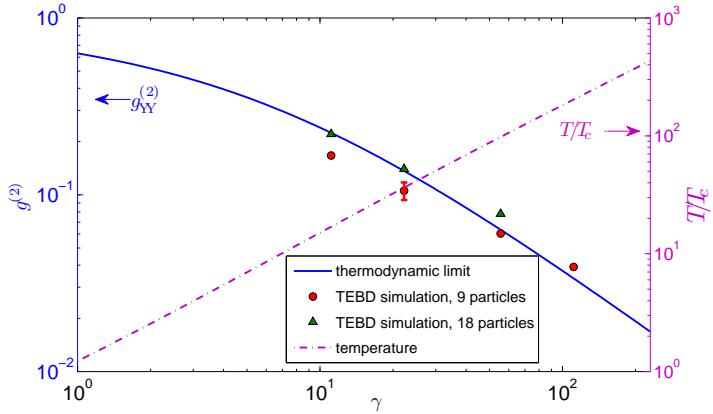,width=0.6\columnwidth}
    \caption{ (Colour online) Steady-state values of $g^{(2)}(0, 0)$ (left scale) for different values
of the interaction parameter $\gamma$ after the interaction quench obtained from TEBD simulations
using 9 (red circles, corresponding to figure \ref{fig:g2-local}) and 18 (green triangles) bosons. Solid line:
value from thermal Gibbs state in the thermodynamic limit; 
dot dashed line: temperature (right scale) corresponding to the given energy per particle in the thermodynamic limit.
 The error bar reflects discretization error estimated by comparing steady-state values for $L=720$ and $L=1440$ lattice
sites as well as error resulting from finite MPS matrix dimension obtained from comparing results for
bond dimension $\chi=100$ and $200$.}
  \label{fig:g2-steady-state}
 \end{figure}

We now want to analyse the local state of the system in the stationary limit. 
In particular we will show that the local steady-state can be well described by the usual finite-temperature 
Gibbs state. To this end we calculate the expected asymptotic value $g^{(2)}_{\rm YY}(0,0)$ from
the thermodynamic Bethe Ansatz \cite{Yang1969}. The system is initially prepared in its 
non-interacting ground state, so we have $g^{(2)}_{\rm ini}(0,0) = 1-1/N$, which in the 
thermodynamic limit $N\to\infty$ approaches unity. The energy of this state 
with respect to the non-interacting Hamiltonian is $0$. At time $t=0$ the
interaction is switched to a finite strength $g>0$ and the expectation value of the 
interaction energy immediately after the quench is given by
\begin{equation}\label{eq:eia0homo}
 \Eia = \int\!\! {\rm d}x\ \frac{g}2 \expv{\left.\hPd\right.^2(x)\hP^2(x)} = \int\!\! {\rm d}x\ \frac{\gamma}{2} 
g^{(2)}(x, x)\rho^3(x).
\end{equation}
Since in a homogeneous system there is no $x$-dependence the energy per particle is
\begin{equation}\label{eq:eia0pphomo}
 \left. \frac{E}{N}\right\vert_{t=0+} = \left.\frac{\Eia}
{N}\right\vert_{t=0+} 
=\gamma\Tc.
\end{equation}
Here we have introduced the critical temperature $\Tc$ in one dimension $\Tc = {\rho^2}/{2}$.
One recognises that the system is in a highly excited non-equilibrium state
after the quench if $\gamma \gtrsim 1$. Using the energy per particle, the density $\rho$
and the Tonks parameter $\gamma$ as input parameter, we can extract a temperature $T$ of a
corresponding thermal Gibbs state by
inverting the Yang-Yang equations of the thermodynamic Bethe Ansatz \cite{Yang1969} for the
excitation-energy and particle densities $\epsilon(q), n(q)$ in momentum space
 \begin{eqnarray*}
\epsilon(\lambda) = \frac{\lambda^2}2 - \mu -\frac{T}{2\pi}\int_{-\infty}^\infty\!\! \! {\rm d}\xi\, K(\lambda,\xi) 
\ln\left(1+{\rm e}^{-\epsilon(\xi)/T}\right),&
\\
2\pi \, n(\lambda) \left(1+{\rm e}^{\epsilon(\lambda)/T}\right)= 1 + \int_{-\infty}^\infty\!\!\! {\rm d}\xi\, K(\lambda,\xi) \, n(\xi).
\end{eqnarray*}
Here $K(\lambda,\xi) = \frac{2g}{g^2+(\lambda-\xi)^2}$ and $\rho =\int{\rm d}\lambda \, n(\lambda)$.
With the help of the Hellmann-Feynman theorem we can then obtain the value $g^{(2)}_{\rm YY}(0, 0)$ corresponding
to the Gibbs state at temperature $T$ \cite{Kheruntsyan2003}:
\begin{eqnarray}\label{eq:g2-HF}
 g^{(2)}_{\rm YY}(0, 0) = \frac{2}{\rho^2} \frac{\partial}{\partial \gamma} f(\gamma,T)
\end{eqnarray}
with $f(\gamma,T)$ being the free energy per particle
\begin{equation}
f(\gamma,T) = \mu - \frac{T}{2\pi \varrho}\int_{-\infty}^\infty\!\! \! {\rm d}\xi\, \ln\left(1+{\rm e}^{-\epsilon(\xi)/T}\right).
\end{equation}
In the limit $1\ll\tau \ll\gamma^2$, where $\tau =T/T_c$, (\ref{eq:g2-HF}) attains the simple form \cite{Kheruntsyan2003}
\begin{equation}
  g^{(2)}_{\rm YY}(0,0) = \frac{2\tau}{\gamma^2}.\label{eq:g2-simple}
\end{equation}
In Figure \ref{fig:g2-steady-state} we have plotted the values of $g^{(2)}_{\rm YY}(0,0)$ from the thermal
Gibbs state in the thermodynamic limit as function of the interaction strength $\gamma$ (solid line). 
Also shown are the steady-state values
obtained from the numerical simulation in Figure \ref{fig:g2-local}. The error bar indicates uncertainties which are here due to discretization artifacts
and error estimates obtained from comparing simulations with MPS bond dimensions $\chi=100$ and $200$. It is available only for one parameter set, since the variation of the discretization length and the bond dimension is numerically expensive. However we expect it to be of about the same relative size for all data points.
One recognises that $g^{(2)}(0, 0; t)$ 
attains in the long-time limit values which are close to that of the thermal
Gibbs state. 
One should note that the steady-state values for the largest values of $\gamma$ ($\gamma = 500/9 $ and 18 bosons as well
as $\gamma = 1000/9$ and 9 bosons) are slightly overestimated in the simulation due to the remaining 
grid artifacts since
here $\langle \hat n\rangle \gamma \approx 0.73$ is no longer small compared to unity.
Also shown is the asymptotic local
temperature of the gas in units of the degeneracy temperature. For large values of $\gamma$, $T\gg T_{\rm c}$,
i.e., after relaxation the gas is in a state with large local temperature.

\section{Non-local Relaxation}

\begin{figure}
   \center\epsfig{file=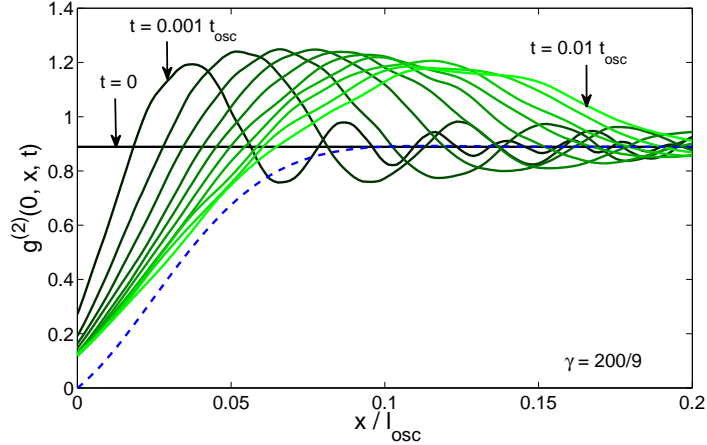,width=0.6\columnwidth}
    \caption{ (Colour online) Time evolution of non-local density-density correlations $g^{(2)}(0, x; t)$ for $\gamma=200/9$. 
$x=0$ denotes the centre of the cloud. One recognises the formation of expanding correlation waves. The dashed blue line
shows the approximation (\ref{eq:g2-approx}) to the non-local correlation in a thermal Gibbs state from \cite{Deuar2009}
multiplied by $g^{(2)}(0,0,t=0)=8/9$ to account for the final particle number ($N=9$) used in the simulation.
}
  \label{fig:g2-non-local}
 \end{figure}

We now discuss the dynamics of non-local quantities. Specifically we consider the non-local two-particle correlation $g^{(2)}(0, x; t)$. In Figure \ref{fig:g2-non-local} we have plotted $g^{(2)}(0, x; t)$ for different times after the interaction quench. One recognises that
while the local correlations attain a steady-state value on a short time scale, the non-local evolution
happens much slower. Switching on the particle-particle repulsion leads to a 
fast reduction of the probability to find two particles at the same position. Associated with this
is a correlation flow to larger distances leading to expanding correlation waves. 
For very short times the propagation velocity of correlation waves is faster than the Fermi
velocity $v_F=\pi \rho$. But at the largest time shown in Figure \ref{fig:g2-non-local} 
corresponding to $t=0.01 t_{\rm osc}$, the maximum of the correlation wave has travelled a distance
of approximately $\Delta x = 0.12 l_{\rm osc}$ which is consistent with the
speed of sound which for large values of $\gamma$ approaches
\begin{equation}
 v_s=v_F \left(1-\frac{4}{\gamma}\right).
\end{equation}

The buildup of a maximum that behaves like a wavefront can be understood as follows: The integral over space of $g^{(2)}(0, x; t)$ is (in a homogeneous system) a constant with respect to time due to particle number conservation. The numerator of $g^{(2)}(0, x; t)$ is proportional to the joint probability distribution to find a particle at position $x$ given that there is one particle at the origin.


So as the quench can not change $g^{(2)}(0, x; t)$ significantly outside the light cone given by the Fermi velocity, the reduction of the probability to find two particles close together must be accompanied by an increase at finite distance.

As one can see from Figure \ref{fig:g2-non-local} also the non-local correlation function 
approaches at least for smaller distances in the large-time limit that of the thermal Gibbs state with temperature and density given by the initial
conditions and the Tonks parameter $\gamma$. For comparison we have plotted an approximation
to the finite-temperature non-local $g^{(2)}$ from reference \cite{Deuar2009} which holds in the regime
$1\ll\tau\ll \gamma^2$
\begin{eqnarray}
g^{(2)}_{\rm T}(0, x) = 1-\left[1-4\sqrt{\frac{\pi\tau}{\gamma^2}}\left(\frac{x}{\lambda_T}\right)\right]
{\rm e}^{-2\pi (x/\lambda_T)^2}.\label{eq:g2-approx}
\end{eqnarray}
Here $\lambda_T=\sqrt{4\pi /\tau \rho^2}$ is the thermal de Broglie length.

\section{Experimental Observation}

 In the following we discuss the possibility to test the local relaxation in
an experiment. For this we make use of the fact that by energy conservation 
the interaction energy lost by the decrease of $g^{(2)}(0, 0)$ must be gained as kinetic energy, $\Ekin(t) = \Eia(t=0) - \Eia(t)$ and the kinetic energy therefore directly gives the value of $g^{(2)}(0, 0, t)$ in the homogeneous case:
\begin{equation}
 \Ekin(t) = \int dx\ \frac{g}{2} \left(1-g^{(2)}(0, 0; t)\right)\rho^2.
\end{equation}
If the interaction is turned on at $t=0$ and turned off abruptly at some time $t=t_1$ the kinetic energy is
the only remaining in the system and can be used to measure $g^{(2)}(0, 0; t_1)$.
\begin{equation}\label{eq:gfromEkinhomo}
 g^{(2)}(0, 0; t_1) = 1-\frac2{\gamma\rho^2}\frac{\Ekin^{\rm final}}{N}.
\end{equation}
In an the experimental setup, the gas must be confined e.g. by an harmonic trapping potential. So the initial 
non-interacting state has a Gaussian density distribution. 
It is also a good assumption, that the correlations decay locally as in the homogeneous 
system corresponding to the local density
provided the density $\rho(x)$ remains constant over the time scale of interest. This is indeed the case, if
$\tint = 1/(\gamma \varrho^2) \ll l_{\rm osc} / v_s \sim l_{\rm osc} /\varrho$
This means, that the Tonks parameter must be large compared to $\frac{1}{N}$, which is of course the case we are interested in. We note that the region in the wings of the density distribution which does not fulfil this constraint gives a negligible contribution to the total interaction energy. Of course measuring the kinetic energy in the trap gives only an average of $\frac{g^{(2)}(x, x)}{\rho^2(x)}$ over the trap.

\section{Summary}

Using the time-evolving block decimation scheme we have numerically analysed the 
dynamics of a 1D Bose gas (LL-model), after an interaction quench from zero to a finite value. 
Although globally the 1D Bose gas does not thermalise, we have shown that local quantities attain 
a steady-state value on a time scale $t_{\rm ia}=(\gamma \rho^2)^{-1}$. Within the achievable 
accuracy these values are consistent with the assumption that local quantities 
relax to a thermal Gibbs state with local temperature determined by the initial energy
and chemical potential. Non-local quantities such as the density-density correlation relax on
a much longer time scale set by the velocity of sound by means of correlation waves propagating out of the sample.

\section*{Acknowledgement}

This work was supported in part by the SFB TR49 of the Deutsche Forschungsgemeinschaft and the
graduate school of excellence MAINZ/MATCOR.


\end{document}